\documentclass[aps,pra]{revtex4}
\usepackage{graphicx}

\begin{document}

\title{Economically Improving Message-Unilaterally-Transmitted Quantum Secure Direct Communication
to Realize Two-Way Communication
\thanks{Email: zhangzj@wipm.ac.cn. }}

\author{Zhan-jun Zhang$^1$, Zhong-xiao Man$^1$ and Yong Li$^2$ \\
{\normalsize $^1$ Wuhan Institute of Physics and Mathematics,
Chinese
Academy of Sciences, Wuhan 430071, China} \\
{\normalsize $^2$ Department of Physics, Central China Normal
University, Wuhan 430079, China} \\ {\normalsize Email:
zhangzj@wipm.ac.cn}}

\date{\today}
\maketitle

\begin{minipage}{380pt}
{\it We present a subtle idea to economically improve
message-unilaterally-transmitted quantum secure direct
communication (QSDC) protocols to realize two-way secure direct communication.} \\

\end{minipage}

Suppose Alice and Bob securely share a secret key. When the
message sender, say, Alice, wants to send her secret messages to
Bob, she can first use the secret key to encrypt her secret
messages, then she publicly sends the encrypted messages to the
message receiver, say, Bob, via a classical channel. After
receiving the encrypted messages, by using the secret key Bob can
decrypt the encrypted messages to securely obtain the secret
messages which Alice wants to send him. According to the classical
one-time-pad method, if the secret key is used one time, then both
the secret messages and the secret key are secure. Quantum key
distribution (QKD) is an ingenious application of quantum
mechanics, in which two remote legitimate users (Alice and Bob)
establish a shared secret key through the transmission of quantum
signals. Hence, much attention has been focused on QKD after the
pioneering work of Bennett and Brassard published in 1984 [1].
Till now there have been many theoretical QKDs [2-20].

Recently, a novel concept, quantum secure direct communication
(QSDC) was proposed [15,16,19]. In those QSDC protocols
[15,16,19], secret messages can be transmitted directly without
first creating a secret key to encrypt them and quantum mechanics
ensures the security of the transmitted secret messages.

Nevertheless, so far all the QSDC protocols [15,16,19,21] are
message-unilaterally-transmitted communication protocols,
alternatively, two legitimate parties can not simultaneously
transmit their different secret messages to each other (dialogue)
in a set of quantum communication device. In general, convenient
bidirectional simultaneous mutual communications (dialogue) are
very useful and usually desired. In fact, the goal to realize a
two-way secure direct communication can be achieved by adopting
the strategy of using two sets of message-unilaterally-transmitted
QSDC device between two parties. Later in this short paper, we
will show the above mentioned strategy of two-way secure direct
communication is uneconomical and we will present a subtle idea to
economically improve a message-unilaterally-transmitted QSDC to
realize two-way secure direct communication.

Suppose that there is a message-unilaterally-transmitted QSDC
device between Alice and Bob and Alice can transmit her secret
messages to Bob via the quantum channel in terms of the QSDC
protocol. The security of Alice's secret message transmission is
ensured by the quantum mechanics in the QSDC protocol. After
Alice's secret message transmission, if both Alice and Bob further
take the secret messages as a shared secret key, then Bob can use
it to encrypt his secret messages and then transmits publicly his
encrypted messages to Alice via a classical channel. Since Alice
knows the so-called secret key, she can extract Bob's secret
messages after receiving Bob's encrypted messages. According to
the classical one-time-pad method, during this transmission via a
classical channel, both the so-called secret key and Bob's secret
messages are secure. Hence in our improved scheme, all the secret
messages can be transmitted securely, either via a quantum channel
or via a classical channel. Moreover, the strategy of message
authentification can be used to protect the secret messages
transmitted either from Alice to Bob via the quantum channel or
form Bob to Alice via the classical channel. To summarize, our
subtle idea is to let both parties take the secret messages
securely transmitted from Alice to Bob via a quantum channel in
terms of a message-unilaterally-transmitted QSDC protocol as their
shared secret key and then Bob directly and securely communicates
with Alice via a public classical channel. By the way, in fact, in
the QSDC protocols [15,16,19,21] classical channel is also
employed. In our improved scheme we only use it more frequently.

So far, one believes that provided that a secret key is securely
shared by two parties in advance the communication cost between
them via a public classical channel is much cheaper than the QSDC
cost via a quantum channel, according to the present-day
technologies. Hence, as mentioned before, the strategy of using
two sets of message-unilaterally-transmitted QSDC device between
two parties to realize a two-way secure communication is
uneconomical. In contrast, our strategy is optimal.

Incidentally, in our previous preprints [22,23], the subtle idea
has also been used without intention. That is, in the works, Bob
introduces the additional unitary operations and at last publicly
announces his measurement outcomes. The essence of his actions is
first to use Alice's secret message as secret key to encrypt his
secret messages and then to communicate with Alice via a public
classical channel. According to our above description of our
subtle idea, one can see that the additional unitary operations in
Refs. 22 and 23 are completely unnecessary. Bob can directly
perform a Bell-state measurement, then he uses the measurement
result as a secret key to encrypt his secret messages.

Although our subtle idea is simple, it is really very important
for it can be applied to any message-unilaterally-transmitted QSDC
protocols to realize two-way secure direct communications. Now
only several QSDC protocols [15,16,19,21] have been proposed and
they are all message-unilaterally-transmitted protocols. Hence,
they all can be improved.

This work is supported by the National Natural Science Foundation
of China under Grant No. 10304022. \\

\newpage
\noindent {\bf References}

\noindent [1] C. H. Bennett and G. Brassard, in {\it Proceedings
of the IEEE International Conference on Computers, Systems and
Signal Processings, Bangalore, India} (IEEE, New York, 1984),
p175.

\noindent[2] A. K. Ekert, Phys. Rev. Lett. {\bf67}, 661 (1991).

\noindent[3] C. H. Bennett, Phys. Rev. Lett. {\bf68}, 3121
 (1992).

\noindent[4] C. H. Bennett, G. Brassard, and N.D. Mermin, Phys.
Rev. Lett. {\bf68}, 557(1992).

\noindent[5] L. Goldenberg and L. Vaidman, Phys. Rev. Lett.
{\bf75}, 1239  (1995).

\noindent[6] B. Huttner, N. Imoto, N. Gisin, and T. Mor, Phys.
Rev. A {\bf51}, 1863 (1995).

\noindent [7] M. Koashi and N. Imoto, Phys. Rev. Lett. {\bf79},
2383 (1997).

\noindent[8] W. Y. Hwang, I. G. Koh, and Y. D. Han, Phys. Lett. A
{\bf244}, 489 (1998).

\noindent[9] P. Xue, C. F. Li, and G. C. Guo,  Phys. Rev. A
{\bf65}, 022317 (2002).

\noindent[10] S. J. D. Phoenix, S. M. Barnett, P. D. Townsend, and
K. J. Blow, J. Mod. Opt. {\bf42}, 1155 (1995).

\noindent[11] H. Bechmann-Pasquinucci and N. Gisin, Phys. Rev. A
{\bf59}, 4238 (1999).

\noindent[12] A. Cabello, Phys. Rev. A {\bf61},052312 (2000);
{\bf64}, 024301 (2001).

\noindent[13] A. Cabello, Phys. Rev. Lett. {\bf85}, 5635 (2000).

\noindent[14] G. P. Guo, C. F. Li, B. S. Shi, J. Li, and G. C.
Guo, Phys. Rev. A {\bf64}, 042301 (2001).

\noindent[15] A. Beige, B. G. Englert, C. Kurtsiefer, and
H.Weinfurter, Acta Phys. Pol. A {\bf101}, 357 (2002).

\noindent[16] Kim Bostrom and Timo Felbinger, Phys. Rev. Lett.
{\bf89}, 187902 (2002).

\noindent[17] G. L. Long and X. S. Liu, Phys. Rev. A {\bf65},
032302 (2002).

\noindent[18] F. G. Deng and G. L. Long, Phys. Rev. A {\bf68},
042315 (2003).

\noindent[19] F. G. Deng, G. L. Long, and X. S. Liu, Phys. Rev. A
{\bf68}, 042317 (2003).

\noindent[20] N. Gisin, G. Ribordy, W. Tittel, and H. Zbinden,
{\it Rev. Mod. Phys.} {\bf 74} 145 (2002).

\noindent[21] F. G. Deng and G. L. Long, {\it Phys. Rev.} A
{\bf69} 052319 (2004).

\noindent[22] Z. J. Zhang, e-print quant-ph/0403186.

\noindent[22] Z. J. Zhang and Z. X. Man, e-print quant-ph/0403215.

\end{document}